\documentclass[traditabstract]{aa}
\usepackage[utf8]{inputenc}
\usepackage{graphicx,amsmath}
\usepackage[english]{babel}
\usepackage{natbib}
\usepackage[T1]{fontenc}
\bibpunct{(}{)}{;}{a}{}{,} 

\begin{document}

\title{Application of the Kolmogorov-Smirnov test to CMB data:\\Is the universe really weakly random?}
\titlerunning{Is the universe really weakly random?}
\author{Sigurd K. Næss}
\institute{Institute of theoretical Astrophysics, University of Oslo, P.O.Box 1029 Blindern, 0315 Oslo, Norway}

\newcommand{\fig}[3]{
	\begin{figure}
	\begin{center}
	\includegraphics[width=0.5\textwidth]{#1}
	\caption{#2}
	\label{#3}
	\end{center}
	\end{figure}
}
\newcommand{\dfig}[4]{
	\begin{figure}
	\begin{center}
	\includegraphics[width=0.45\textwidth]{#1}
	\includegraphics[width=0.45\textwidth]{#2}
	\caption{#3}
	\label{#4}
	\end{center}
	\end{figure}
}
\newcommand{\dhfig}[4]{
	\begin{figure}
	\begin{center}
	\begin{tabular}{cc}
	\includegraphics[width=0.22\textwidth]{#1} &
	\includegraphics[width=0.22\textwidth]{#2}
	\end{tabular}
	\caption{#3}
	\label{#4}
	\end{center}
	\end{figure}
}

\abstract{
  A recent application of the Kolmogorov-Smirnov test to the WMAP 7
  year W-band maps claims evidence that the CMB is ``weakly random'',
  and that only 20\% of the signal can be explained as a random
  Gaussian field. I here repeat this analysis, and in contrast to the
  original result find no evidence for deviation from the standard
  $\Lambda$CDM model. Instead, the results of the original analysis
  are consistent with not properly taking into account the
  correlations of the $\Lambda$CDM power spectrum.}


\maketitle

\section{Introduction}
In astronomical data analysis, it is often useful to be
able to test whether a set of data points follows a given
distribution or not. For example, many analysis techniques
depend on instrument noise being Gaussian, and to avoid
bias, one must check that this actually is the case. There
are many different ways in which two distributions can differ,
and correspondingly many different ways to test them for
equality. The simplest ones, such as comparing the means
or variances of the distributions, suffer from the problem
that there are many ways in which distributions can differ
that they cannot detect no matter how many samples are available.
For example, samples from a uniform distribution can easily
pass as Gaussian if one only considers the mean and variance.

The popular Kolmogorov-Smirnov test (K-S test) resolves this problem by
considering the cumulative distribution functions (CDF) instead:
Construct the empirical CDF of the data points and
find its maximum absolute difference $K$ from the theoretical CDF.
Due to the limited number of samples, the empirical CDF will
be noisy, and $K$ will therefore be a random variable with
its own CDF, which in the limit where the number of samples
goes to infinity is given by
\begin{align}
P(x<K) = F_\mathrm{KS}(\sqrt{N_\mathrm{obs}} K) \label{KS}
\end{align}
with
\begin{align}
F_{KS}(x) = 1-2\sum_{i=1}^\infty (-1)^{i-1}\mathrm{e}^{-2i^2x^2}.
\end{align}
In contrast with the simplest tests, this test can detect
any deviation in the distributions, but may require a
large number of samples to do so, especially in the tails
of the distribution.

Recently, a series of papers \citep{
2011A&A...525L...7G,2008A&A...492L..33G,2010EL.....9119001G} has applied
this test to WMAP's cosmic microwave background (CMB) maps,
resulting in the remarkable claim that the CMB is
``weakly random'', with only 20\% of the CMB signal behaving
as one would expect from a random Gaussian field. This result
went on to be used in a much discussed series of papers
\citep{2010arXiv1011.3706G,2010arXiv1012.1486G,2011arXiv1104.5675G}
claiming a strong detection of concentric low-variance circles in
the CMB, which was taken as evidence for Conformal Cyclic Cosmology.
Other groups failed to significantly detect the circles
\citep{2010arXiv1012.1268W,2011JCAP...04..033M,2010arXiv1012.1656H}.
The difference in significance was due different CMB models:
\cite{2010arXiv1012.1268W,2011JCAP...04..033M,2010arXiv1012.1656H}
used realizations of the best-fit $\Lambda$CDM power
spectrum, while \cite{2010arXiv1012.1486G} used a ``weakly random'' CMB model.

Both in order to resolve this issue, and because a weakly random
universe would be a strong blow against the $\Lambda$CDM
model in its own right, it is important to test this result.

\section{Method}
Before applying the K-S test, one must be aware of its limited
area of validity: Equation~(\ref{KS}) requires an infinite
number of independently identically distributed samples, while
CMB maps actually consist of a limited number of correlated
samples. However, both the correlations and number of samples
can be compensated for, as we shall see.

\subsection{Application of the K-S test to correlated data}
\label{sect_corr}
Though the K-S test is not immediately applicable to a correlated
data set, it is possible to perform an equivalent test on
a transformed set of samples. The question we are trying to answer
with the K-S test is ``Do the samples follow
the theoretical distribution?''. The truth or falseness of this
is preserved if we apply the same transformation to both the samples
and the distribution we test them against, and to be able to use the
K-S test, the logical transformation to use is a \emph{whitening}
transformation, which results in an independent, identical distribution
for the samples.  With
original samples $\bf d$ with covariance matrix $\bf C$, the whitened
(uncorrelated with unit variance) samples $\bf r$ are given by:
\begin{align}
\bf r = \bf C^{-\frac12}\bf d \label{white}
\end{align}
Thus, to test whether the data points $\bf d \leftarrow N(0,\bf C)$,
we can test the equivalent hypothesis $\bf r \leftarrow N(0,1)$.

In the case of CMB maps, both the data itself and the noise is expected
to be Gaussian, so the obvious theoretical distribution here is
$N(0,\bf S + \bf N)$,
where the CMB signal covariance matrix $\bf S$ is given by the
two-point correlation function:
\begin{align}
S_{ij} = \sum_{l=0}^\infty \sqrt{\frac{2l+1}{4\pi}}C_l B_l P_l(\cos(|\bf p_i-\bf p_j|))
\end{align}
$P_l(x)$ are the Legendre polynomials normalized to $\frac{1}{2\pi}$,
and $\bf p_i$ and $\bf p_j$ are the direction vectors for pixel $i$
and $j$ in the disk. $C_l$ is the $\Lambda$CDM angular power spectrum,
while $B_l$ accounts for the beam and pixel window.
$\bf N$ is instrument dependent, but for the WMAP W-band
CMB map we will use here, the noise is nearly diagonal, and given by the
corresponding W-band RMS map.
\fig{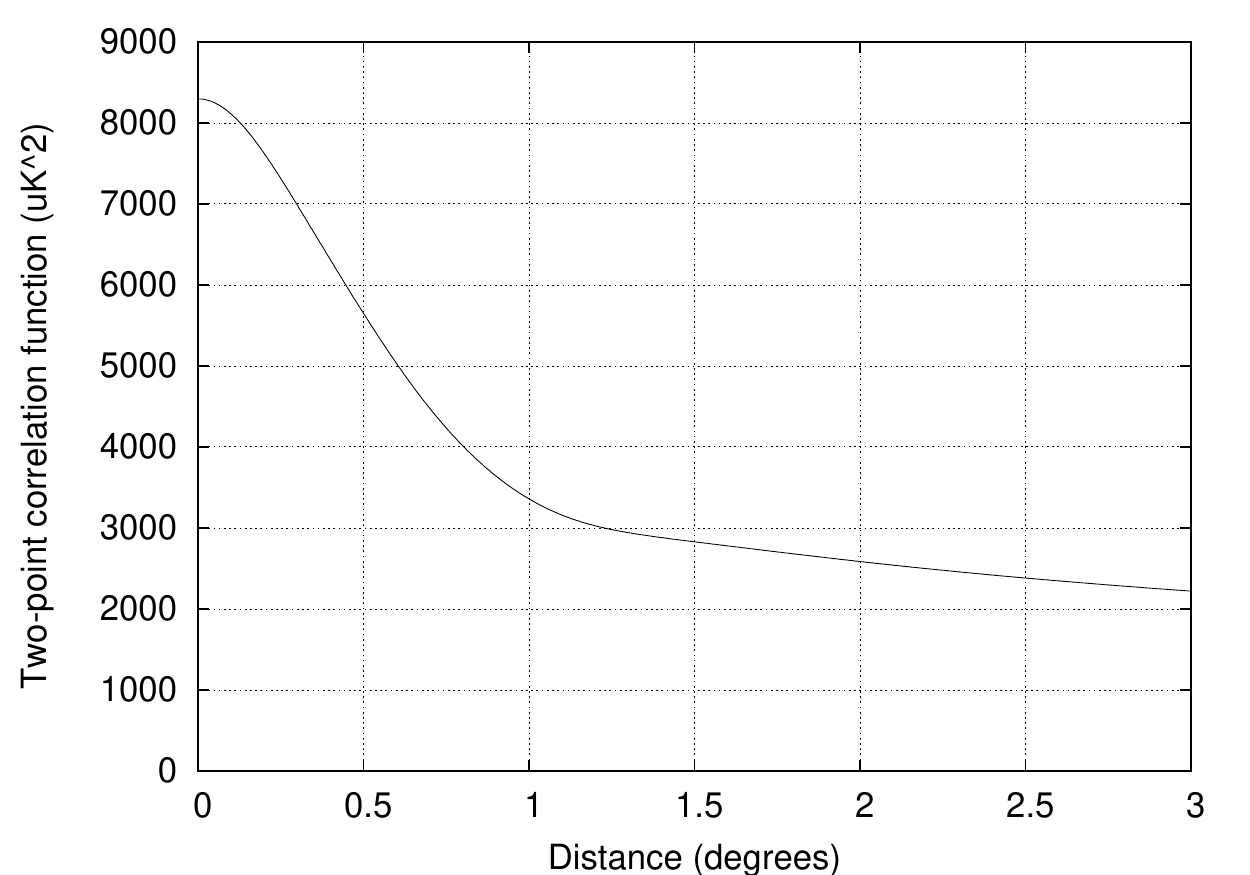}{$\Lambda$CDM two point correlation function
after applying the WMAP W-band beam and the HEALPix \citep{2005ApJ...622..759G}
nside 512 pixel window.}{corrfun}

\subsection{Application of the K-S test with few samples}
The other problem we need to account for is our finite number
of samples. In this case
equation~(\ref{KS}) is only approximate. For most uses of the
test, this approximation is good enough, especially when
employing analytical expressions for improving the quality of
the approximation for low numbers of samples
\citep{1964mtps.book.....V}. For example, when performing a single
test to accept or reject a test distribution, a bias of a few percent
in the confidence with which the hypothesis is rejected is not important.

However, when making statistics for a large number of such test results,
such a bias may make the results ambiguous. Given a set of experiments
with a corresponding set of maximum deviations $\{K_i\}$, the corresponding
probabilities $\{p_i = P(x<K_i)\}$ should be uniformly distributed if the samples
actually follow the theoretical distribution, and a histogram of $\{p_i\}$
should therefore be flat. Deviations from this indicate that the theoretical
distribution does not accurately describe the samples. However, the approximate
equation~(\ref{KS}) also introduces a small non-uniformity in $\{p_i\}$ even
if the samples actually do follow the distribution. To avoid the
ambiguity this causes, we will instead compute a numerical correction function
mapping the approximate $p$ to the true $p'$.
\footnote{
What we do here is essentially replacing the analytical
Komolgorov distribution (equation (\ref{KS})) with a numerical
distribution. This could also be done without using the analytical
distribution as a basis, at a small cost in clarity.}

To build up the correction, we simulate a large number\footnote{
The number necessary depends on the level of accuracy desired.
The noise in the estimate of $G(p)$ propagates to the final
results. To make this a subdominant noise contribution, the number
of simulations should be at least as large as the number of
actual experiments, preferably much higher.} of experiments,
each with the same number of samples as the actual data set, but drawn
directly from the theoretical distribution. Thus, for these, $\{p_i\}$
should be uniform, with a CDF of $G_\infty(p) = p$. However, since
equation~(\ref{KS}) is inexact, for small numbers of samples, the actual
CDF is $G(p) \ne G_\infty(p)$. The mapping between the approximate $p$
and true $p'$ is given by $G(p) = G_\infty(p')
\Rightarrow p' = G_\infty^{-1}(G(p)) = G(p)$. Thus, for a limited number
of samples
\begin{align}
P(x<K) = G(F_\mathrm{KS}(\sqrt N_\mathrm{obs} K)). \label{KS2}
\end{align}
Figure~\ref{fig_kscorr} illustrates the correction function for
$5\cdot 10^6$ simulations of 540 each. For this many samples, the
correction is only of the order of 1\%.
\fig{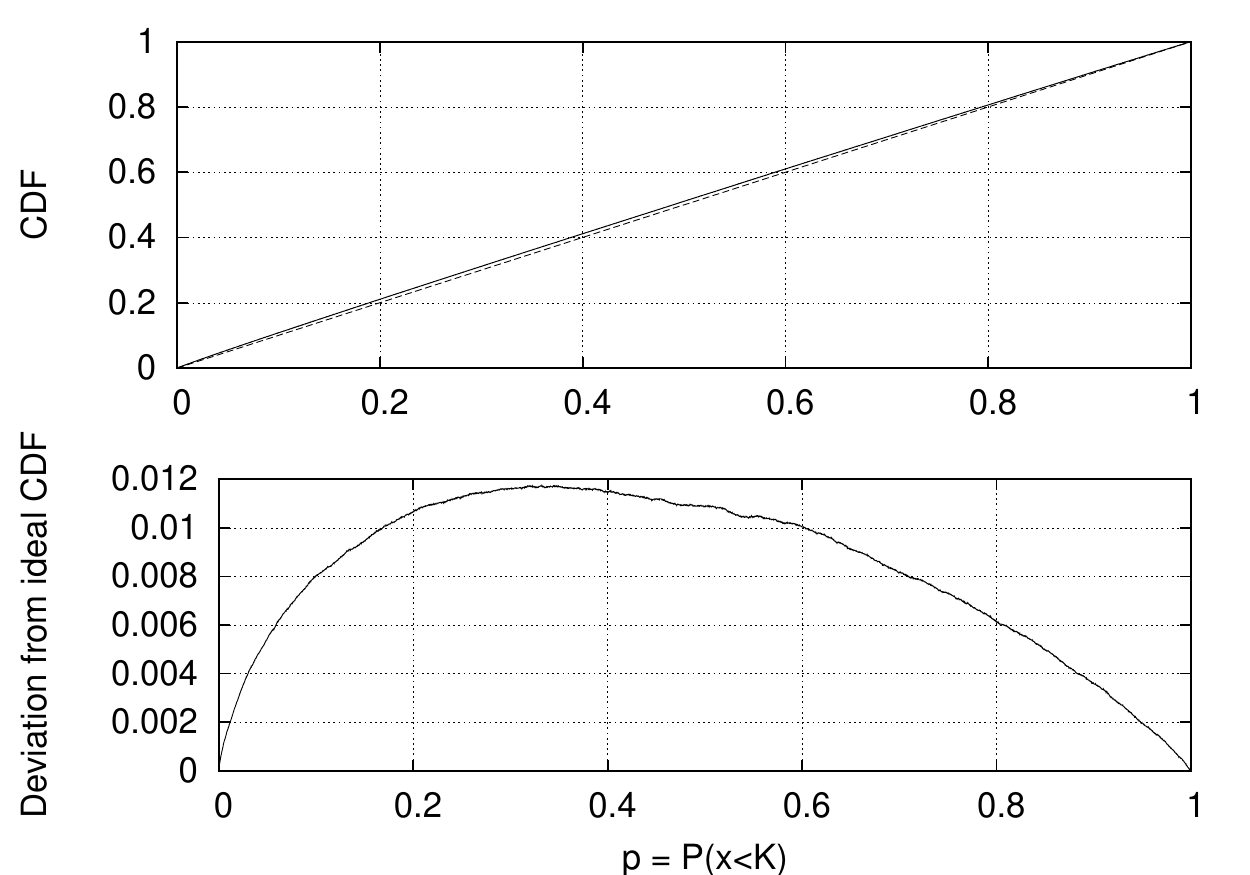}{When applying the K-S test to samples known to
come from the correct distribution, the resulting values $\{p_i=P(x<K_i)\}$
should be uniformly distributed, but when working with a limited number
of samples, the Kolmogorov distribution is only approximate, and the
actual CDF of the results, $G(p)$, differs from the ideal $G_\infty(p)=p$.
This is shown in the upper panel for the case of 540 samples per experiment,
where $G(p)$ is the solid line and $G_\infty(p)$ is dashed.
The lower panel shows the deviation between the two, which is of the order of
1\% in this case (but larger with fewer samples).}{fig_kscorr}

\section{Does $\Lambda$CDM fail the K-S test?}
\label{results1}
With this in hand we can finally apply the K-S test on CMB data. Following
\cite{2011A&A...525L...7G}, we randomly pick 10\,000 disks with
a radius of 1.5 degrees from the WMAP 7 year W-band map \citep{2011ApJS..192...14J},
with the region within 30 degrees from the galactic equator excluded.
Each disk contains on average 540 pixels, which are whitened using
equation~(\ref{white}). A typical disk before and after the whitening operation
can be seen in Fig.~\ref{fig_disks}. After whitening, the values
should follow the distribution $N(0,1)$ if our model is correct.
\dhfig{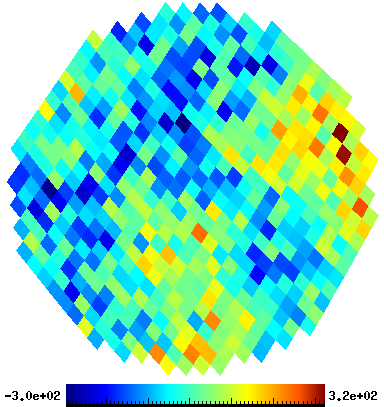}{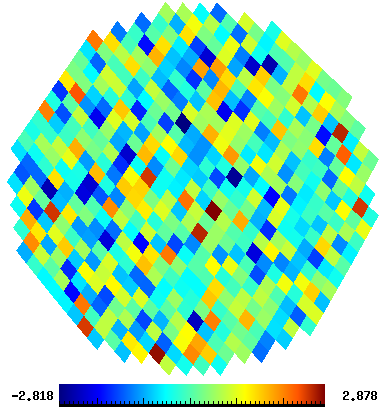}{A randomly selected disk
before (left) and after (right) the whitening operation.
The samples are strongly correlated and thus unsuitable for the
K-S test before the transformation, but afterwards no correlations
are visible and the variance is 1. Note that whitening the data does
not mean that we are ``forcing'' the K-S test to pass. The whitened
data will only end up matching $N(0,1)$ after whitening if they followed
our theoretical distribution $N(0,\bf C)$ before.}{fig_disks}

The histogram of resulting probabilities $\{p_i = P(x<K_i)\}$ from
of applying equation~(\ref{KS2}) to the hypothesis $\bf r \leftarrow N(0,1)$
is shown in Fig.~\ref{fig_ks_result}, together with the 68\% and 95\%
intervals from 300 simulations. The data and simulations are consistent,
and follow a uniform distribution as expected\footnote{
It should be noted that the histogram bins are not completely
independent for two reasons: Firstly, some disks are going to overlap, meaning
that the same samples enter into several different K-S tests, and secondly,
while our transformation has made the samples within each disk independent,
the correlation between different disks is still present.}: The CMB
map is fully consistent with $\Lambda$CDM + WMAP noise as far as the K-S
test is concerned.

This is dramatically different from the
curve found by \cite{2011A&A...525L...7G}, which was strongly biased towards \emph{low}
values. Low values of $P(x<K)$ would mean that the empirical CDF of the
samples matches the theoretical one \emph{too well}, i.e. even better
than samples drawn directly from the theoretical distribution.
\dfig{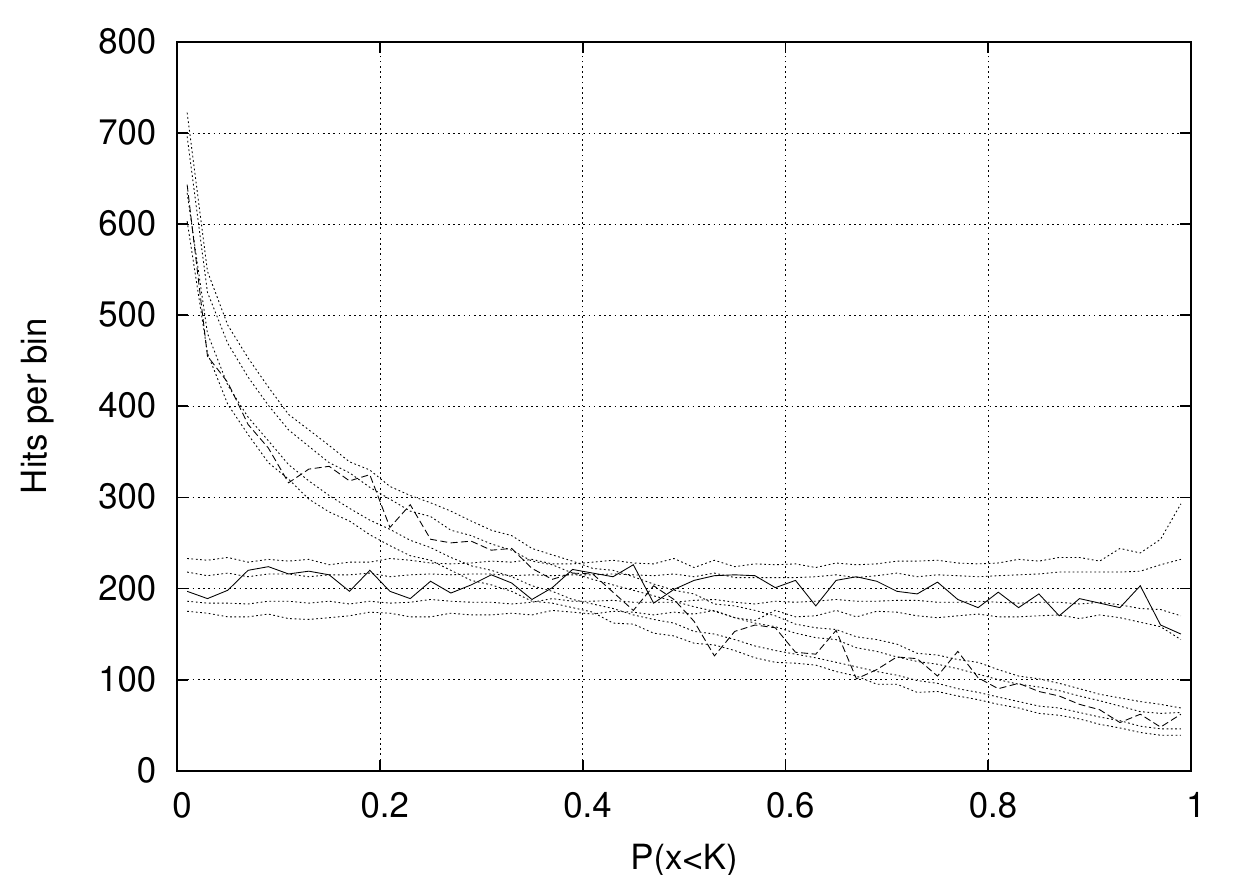}{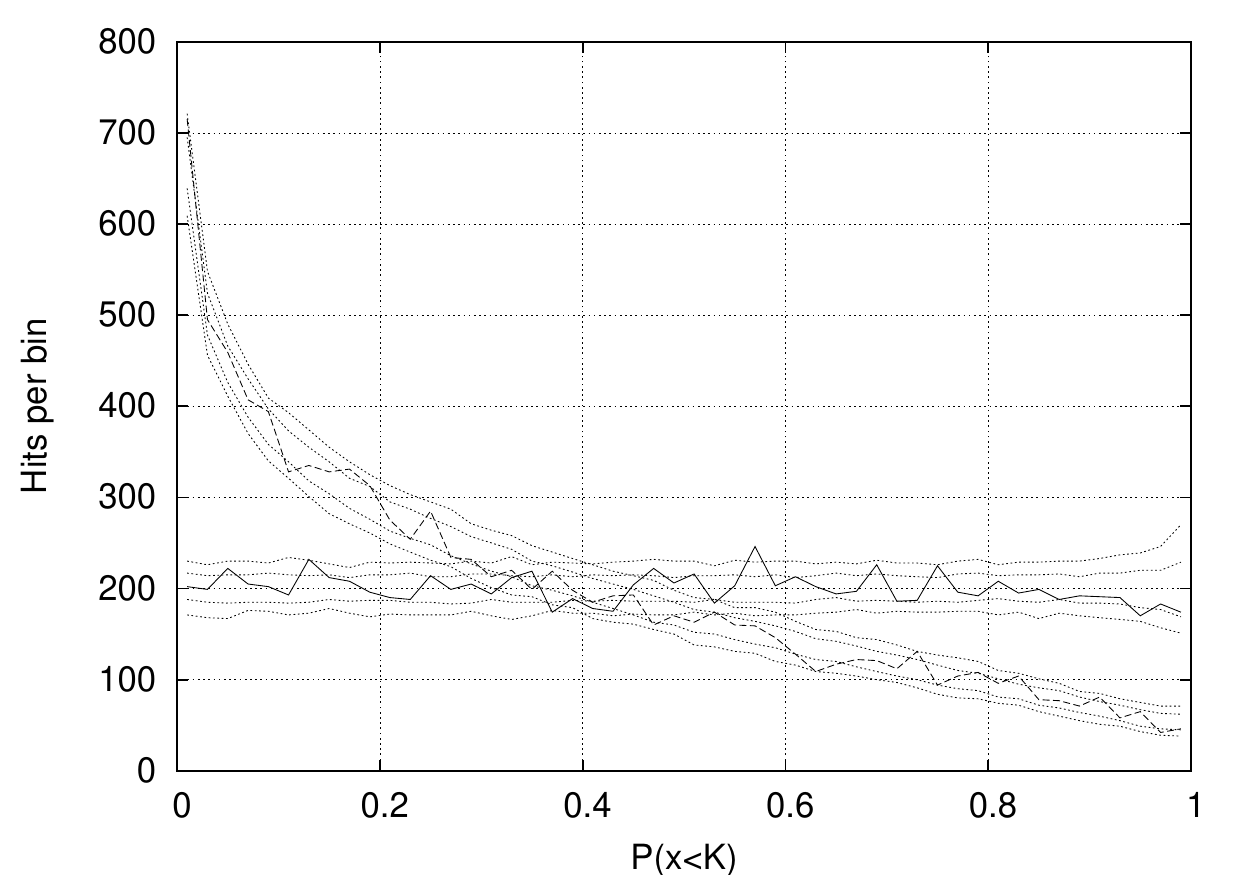}{Histogram of results of the K-S test.
Each panel compares the results from
properly taking the correlations into account (solid line) with those one gets from
ignoring them (dashed line), together with 68\% and 95\% intervals (dotted lines)
from simulations. The upper panel corresponds to using samples
further than 30 degrees away from the galactic equator, while the lower
panel instead uses the WMAP KQ85 analysis mask. In both cases, both the
map and the simulations pass the K-S test when taking the correlations
into account, while if the are ignored, the K-S test fails in the same way
\cite{2011A&A...525L...7G} reported.}{fig_ks_result}

What could cause \citeauthor{2011A&A...525L...7G} to get results so different from ours?
One way biasing $P(x<K)$ low is by basing the parameters of your
test distribution on the values themselves. However, even without
doing this, it is possible to get low values if the values used
in the K-S test are \emph{correlated}.
This is also consistent with the presentation given by
\cite{2011A&A...525L...7G} who apparently applied the K-S test directly to the raw
samples $\bf d$,
or equivalently, that they model the pixel values as coming from a
1-dimensional distribution. To check this, I repeated the analysis,
this time using the theoretical distribution $d \leftarrow N(\mu,\sigma^2)$,
where $\mu$ and $\sigma^2$ are the measured mean and variance of the
samples in the disk. The result is also shown in Fig.~\ref{fig_ks_result}.
This time, the bias towards low values is clearly recreated.

It therefore seems likely that \citeauthor{2011A&A...525L...7G}'s
reported ``weak randomness''
is the result of not properly taking the CMB's correlations into account.
One is, of course, free to use whatever distribution one wants as the
theoretical distribution in a K-S test, even a model where the CMB
pixels are independently identically distributed, with no correlations
at all. The problem lies in the interpretation of the test
results. For \cite{2011A&A...525L...7G}, the K-S test results are clearly
not uniform, indicating
that the chosen theoretical distribution has been disproven. However,
\citeauthor{2011A&A...525L...7G} then go on to create a set of simulations (linear combinations
of 20\% Gaussian and 80\% static signal) that fail the test in the same way
as the WMAP map does. But having two sets of samples fail the K-S test the
same way does not prove that they have the same properties. It simply means
that the chosen test distribution was a poor choice.

\section{Kolmogorov maps}
\label{results2}
While \citeauthor{2011A&A...525L...7G}'s Kolmogorov statistics
are biased by not taking the correlations into account, the
approach of making sky maps of K-S test results introduced in
\cite{2009A&A...497..343G} is still an interesting way to
search for regions of the sky that do not follow the expected
distribution. Making an unbiased Kolmogorov map straightforwardly
follows the procedure in Sect.~\ref{results1}, with the
main difference being the selection of pixels. Instead of
randomly selecting disks, we now
systematically go through nside 16 pixels, using the
1024 nside 512 subpixels inside each one as the samples. These are
then tested against $N(0,\bf C)$ by whitening them via equation~(\ref{white})
and then comparing the whitened samples to $N(0,1)$.

The result
is the nside 16 map of $P(x<K)$ shown in Fig.~\ref{fig_kolmogorov_map}.
Regions that pass the test have a value uniformly distributed between 0 and 1,
and we see that this applies to the CMB-dominated areas of the sky,
while areas dominated by the galaxy fail the test as expected.

For comparison, Fig.~\ref{fig_kolmogorov_map} also includes
the result of making the same
map while ignoring correlations. In this case, the whole sky fails the
test: The CMB-dominated areas are biased low, while the galaxy is
biased high. This map is similar to the map in \cite{2009A&A...497..343G},
which is also too low outside the galaxy, and too high inside, which is,
again, consistent with \citeauthor{2009A&A...497..343G} applying the
K-S test directly to the raw samples.

\dfig{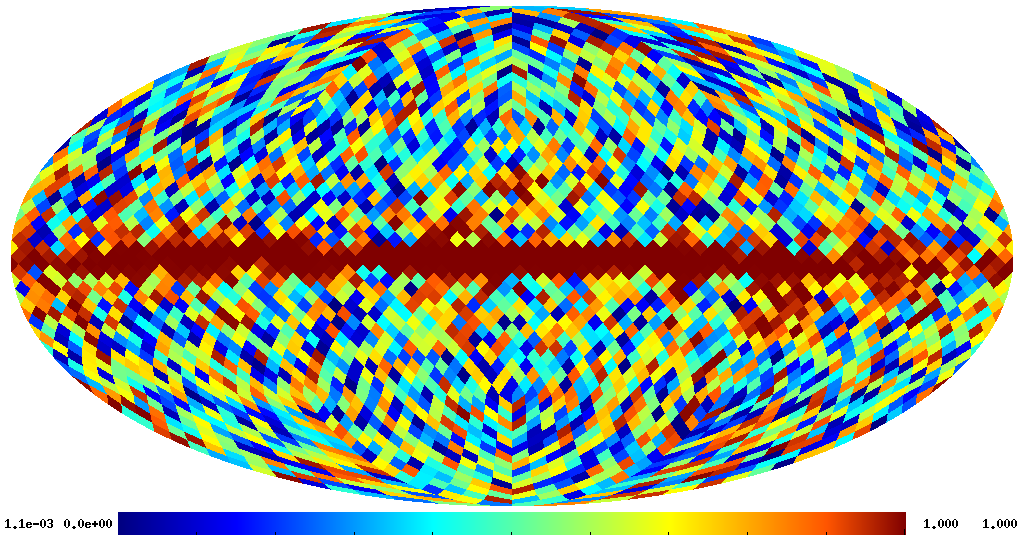}{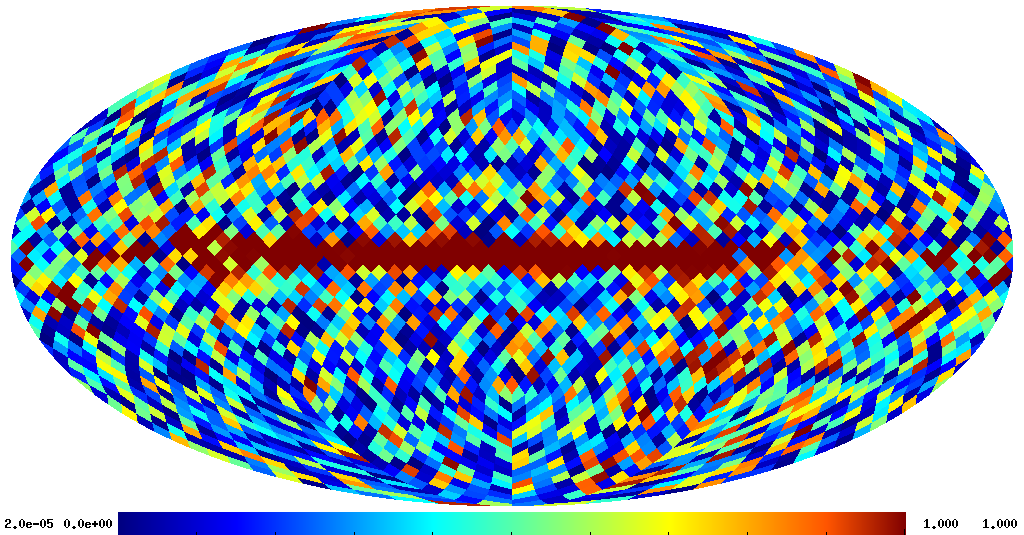}{nside 16 map of $P(x<K)$ based on
pixels from the WMAP 7 year nside 512 W-band map. This time, the K-S test
is performed on the 1024 nside 512 pixels inside each nside 16 pixel
instead of on disks. The upper panel uses $N(0,\bf S+ \bf N)$
as the theoretical distribution (by testing the whitened data against
$N(0,1)$), while the lower panel ignores the correlations, instead
using $N(\mu,\sigma^2)$, where $\mu$ and $\sigma$ are the measured mean
and standard deviation of the samples. The former passes the test outside
the galactic plane, while the latter fails everywhere, being biased low
outside the galaxy.}{fig_kolmogorov_map}

\section{Summary}
The Kolmogorov-Smirnov test is a useful and general way of testing
whether a data set follows a given distribution or not. However,
it only applies to independently identically distributed samples.
The CMB is strongly correlated, and thus not immediately compatible
with the test. However, this can be resolved by the application of
a whitening transformation, replacing the hypothesis $\bf d \leftarrow
N(0,\bf C)$ with the equivalent $\bf C^{-\frac12}\bf d \leftarrow N(0,1)$.
With this, we find that the $\Lambda$CDM passes the K-S test. This is
incompatible with the original
analysis by \cite{2011A&A...525L...7G}, which claimed detection of an
unknown non-random component making up 80\% of the CMB based on
the CMB failing the K-S test there. It turns out that this
analysis did not take the CMB correlations into account, which
we confirm by producing the same failure of the K-S test when
we skip the whitening step. When the correlations are handled
properly, there is no need for a weakly random universe.

\begin{acknowledgements}
The author would like to thank Hans Kristian Eriksen for useful discussion
and comments.
\end{acknowledgements}

\bibliographystyle{aa}
\bibliography{refs.bib}

\end{document}